\documentclass[aps,prd,onecolumn,groupedaddress,showpacs,nofootinbib,amssymb
]{revtex4}
\usepackage[dvips]{graphicx}
\usepackage{amssymb}
\usepackage{amsmath}
\usepackage{graphicx}
\usepackage{amsfonts}
\usepackage{bm}

\begin{document}

\title{Effects of Finite-time Singularities on Gravitational Waves}
\author{K. Kleidis,$^{1}$} \author{V.K.~Oikonomou,$^{1}$}
\affiliation{ $^{1)}$Department of Mechanical Engineering\\ Technological
Education Institute of Central Macedonia \\
62124 Serres, Greece
}

\begin{abstract}
We analyze the impact of finite-time singularities on gravitational waves, in the context of $F(R)$ gravity.
We investigate which singularities are allowed to occur during the inflationary era, when gravitational waves are considered, and we discuss the quantitative implications of each allowed singularity. As we show, only a pressure singularity, the so-called Type II and also a Type IV singularity are allowed to occur during the inflationary era. In the case of a Type II, the resulting amplitude of the gravitational wave is zero or almost zero, hence this pressure singularity has a significant impact on the primordial gravitational waves. The case of a Type IV singularity is more interesting since as we show, the singularity has no effect on the amplitude of the gravitational waves. Therefore, this result combined with the fact that the Type IV singularity affects only the dynamics of inflation, leads to the conclusion that the Universe passes smoothly through a Type IV singularity.
\end{abstract}

\maketitle

\section{Introduction}

Finite-time singularities occur quite frequently in modified gravity theoretical frameworks, and these vary from soft singularities, like the Type IV singularity, to crushing type singularities, like the Big Rip. Finite-time singularities in cosmology where firstly classified in Ref. \cite{Nojiri:2005sx}, and according to the classification of \cite{Nojiri:2005sx}, there are four different types of finite-time singularities. For the classification, the authors of  \cite{Nojiri:2005sx}, used four physical quantities which may or may not blow up at the time instance that the singularity occurs. On the other hand, timelike singularities occur quite frequently in $F(R)$ theories of gravity \cite{fr1,fr2,fr3,fr4,fr5,fr6,fr7,fr8}. In this paper we shall investigate the effects of timelike singularities on the amplitude of gravitational waves which are produced by an $R^2$ theory of gravity. We will be interested in the inflationary era \cite{inflation1,inflation2}, \cite{inflation3,inflation3a},  \cite{inflation4}, \cite{inflation5},  \cite{inflation6},  \cite{inflation7}, since for $F(R)$ theories, an excess of gravitational waves during the inflationary era is known to exist. Particularly, we shall find the Hubble rate which can incorporate a finite-time singularity and then we shall calculate the corresponding amplitude of the gravitational wave. As we will show, only the so-called Type IV singularity has no observable effect on the amplitude of the gravitational wave modes. With regard to the Type IV singularity, the fact that this does not affect the generation and the evolution of the gravitational waves, makes these singularities interesting from a phenomenological point of view, since the Universe seems to pass smoothly through these. For similar recent results on the Type IV singularities, see \cite{sergeioikonomou1,sergeioikonomou2}, where it was shown that the Type IV singularities do not affect the observational indices during inflation, but only affect the dynamics of inflation.

This paper is organized as follows: In the section II we shall present in brief all the essential information for finite-time singularities. In section III we shall calculate the effect of all the known finite-time singularities on the amplitude of the gravitational waves and finally the conclusions follow in the end of the paper.

Before we proceed, let us briefly present the geometric conventions we shall use in this paper. We assume that the geometric background is that of a flat Friedmann-Robertson-Walker (FRW), with line element,
\begin{equation}
\label{metricfrw} ds^2 = - dt^2 + a(t)^2 \sum_{i=1,2,3}
\left(dx^i\right)^2\, ,
\end{equation}
where $a(t)$ stands for the scale factor of the Universe. Moreover, the Ricci scalar for the FRW metric (\ref{metricfrw}) is,
\begin{equation}
\label{ricciscal}
R=6(2H^2+\dot{H})\, ,
\end{equation}
with $H$ standing for the Hubble rate of the Universe, which is $H=\dot{a}/a$.

\section{Gravitational Waves and Finite-time Singularities}

Finite-time singularities were classified for the first time in Ref. \cite{Nojiri:2005sx}, where the authors used four physical quantities for the classification, namely, the total effective energy density $\rho$, the total effective pressure $p$, the scale factor $a(t)$ and the Hubble rate $H(t)$. According to the classification of Ref. \cite{Nojiri:2005sx}, there are four different types of finite-time singularities, which are the following:
\begin{itemize}
\item Type I (``The Big Rip Singularity''): The Big Rip singularity is the most ``severe'' in comparison to all other timelike singularities, from a phenomenological point of view. In particular, if the singularity occurs at a time instance $t=t_s$, then the scale factor $a$, the effective energy density $\rho_{\mathrm{eff}}$ and the pressure $p_{\mathrm{eff}}$, diverge as $t\to t_s$, that is, $a \to \infty$, $\left|p_{\mathrm{eff}}\right| \to
\infty$ and $\rho_{\mathrm{eff}} \to \infty$.
\item Type II (``The Sudden Singularity''): In this case, the scale factor $a(t)$ and the total effective energy density $\rho_{\mathrm{eff}}$ are finite, but the effective pressure $p_{\mathrm{eff}}$ diverges as $t\to t_s$, that is, $a\to a_s$, $\rho_{\mathrm{eff}}\to \rho_s$ but $\left|p_{\mathrm{eff}}\right| \to \infty$.
\item Type III: In this case, only the scale factor is finite, and the effective pressure and effective density diverge at $t\to t_s$, that is, $\left|p_{\mathrm{eff}}\right| \to \infty$, $\rho_{\mathrm{eff}} \to \infty$, but $a \to a_s$.
\item Type IV: This singularity is the most phenomenologically harmless, and the Universe passes smoothly through this. In this case, the scale factor, the effective pressure and the effective density are finite at $t\to t_s$. Also the Hubble rate and it's first derivative are also finite, but the higher derivatives of the Hubble rate diverge at $t\to t_s$.
\end{itemize}
In the next section we shall take into account all the above features and we shall study the effect of the singularities on the amplitude of the gravitational waves.

\section{Gravitational Waves and Finite-time Singularities}

For our gravitational waves study we shall consider the $R^2$ inflation model \cite{starobinsky}, a singular version of which was studied in \\ Ref. \cite{sergeioikonomoujord} for the case of a Type IV singularity. We will consider a vacuum $F(R)$ gravity, with Jordan frame action,
\begin{equation}\label{jordanfrvaccumfr}
S=\frac{1}{16\pi G}\int \mathrm{d}^4x\sqrt{-g}\,F(R)\, ,
\end{equation}
so we assume that no matter fluids are present. In the case of the $R^2$ inflation, the $F(R)$ gravity function has the following functional form,
\begin{equation}\label{starobfr}
F(R)=R+\frac{1}{6M^2}R^2\, ,
\end{equation}
with the parameter $M$ satisfying $M\gg 1$. For the action of Eq. (\ref{jordanfrvaccumfr}), the FRW equations have the following form,
\begin{align}\label{frwequationsfull}
& 6F'H^2=F'R-F-6H\dot{F'}, \\ \notag &
-2F'\dot{H}=\ddot{F}'-H\dot{F}'\, ,
\end{align}
where the ``prime'' in the above equations denotes differentiation with respect to the Ricci scalar, that is $F'=\frac{\mathrm{d}F(R)}{\mathrm{d}R}$. For the $R^2$ gravity of Eq. (\ref{starobfr}), the FRW equations are,
\begin{align}\label{takestwo}
& \ddot{H}-\frac{\dot{H}^2}{2H}+\frac{M^2}{2}H=-3H\dot{H},\\ \notag &
\ddot{R}+3H\dot{R}+M^2R=0\, ,
\end{align}
where $R$ is the Ricci scalar, which in our case is given in Eq. (\ref{ricciscal}).
During a slow-roll phase, the terms $\ddot{H}$ and $\dot{H}$ can be omitted, and the solutions of the resulting differential equations (\ref{takestwo}) are,
\begin{equation}\label{hubstar}
H(t)\simeq H_i-\frac{M^2}{6}\left ( t-t_i\right )\, ,
\end{equation}
\begin{equation}\label{hubstarricciscal}
R(t)\simeq 12 H(t)^2-M^2\, ,
\end{equation}
with $t_i$ being the time instance that inflation is assumed to start, and in addition $H_i$ is the value
of the Hubble rate at the time instance $t=t_i$. Notice that the above solution (\ref{hubstar}) can easily be obtained by the first FRW equation in (\ref{takestwo}), since if we omit the terms $\ddot{H}$ and $\dot{H}$, then the resulting differential equation is $\dot{H}\simeq -M^2/6$, which has the solution (\ref{hubstar}). Also the solution (\ref{hubstarricciscal}) can be obtained since $\dot{H}\simeq -M^2/6$ from the solution (\ref{hubstar}) and by substituting  in the expression for the Ricci scalar (\ref{ricciscal}), we get the expression of Eq. (\ref{hubstarricciscal}). Consider now the following modification of the solution (\ref{hubstar}),
\begin{equation}\label{singstarobhub}
H(t)\simeq H_i-\frac{M^2}{6}\left ( t-t_i\right )+f_0\left (t-t_s
\right)^{\alpha} \, .
\end{equation}
This modification of the Hubble rate leaves room for a singularity to occur at $t=t_s$, depending on the values of the parameter $\alpha$. If we assume that $H_i\gg f_0$, $M\gg f_0$ and in addition that $f_0\ll 1$, the term that can contain the singularity $\sim \left (t-t_s
\right)^{\alpha}$ is much more smaller than the rest of the terms in (\ref{singstarobhub}), therefore for cosmic times near $t=t_s$, the $F(R)$ gravity that can approximately produce the evolution (\ref{singstarobhub}) is again the $R^2$ model of Eq. (\ref{starobfr}), if of course the singularity term tends to zero as $t\to t_s$. This can only happen if $\alpha >0$ and this will be a crucial assumption for the analysis of the amplitudes of the gravitational waves that we develop later on in this section. Now the question is what is the singularity structure of the cosmological evolution (\ref{singstarobhub}), depending on the values of the parameter $\alpha$. As it can be easily shown by deriving the effective energy density and pressure from the $F(R)$ gravity equations, the singularity structure of the evolution (\ref{singstarobhub}) depending on the values of the parameter $\alpha$ is as follows,
\begin{itemize}\label{lista}
\item For $\alpha<-1$ a Big Rip singularity occurs.
\item For $-1<\alpha<0$ a Type III singularity occurs.
\item For $0<\alpha<1$ a Type II singularity occurs.
\item For $\alpha>1$ a Type IV singularity occurs.
\end{itemize}
Obviously the requirement $\alpha>0$ excludes the possibility that a Big Rip and a Type III singularity occur, so what remains is to have a Type II or a Type IV singularity during the inflationary era, since for these two singularities, the Hubble rate (\ref{singstarobhub}) is generated approximately by the $R^2$ gravity of Eq. (\ref{starobfr}), at least for cosmic times near the time instance $t=t_s$. Hence the focus in the rest of this paper is to find the effect of a Type II and of a Type IV singularity on the amplitude of a gravitational wave. Notice that the Type II singularity is a pressure singularity, so from the inflationary dynamics point of view, it is rather unusual, if not impossible, to occur during the inflationary era, as was discussed in \cite{sergeioikonomoujord}, but one of the tasks of this work is to examine the impact of this singularity on the amplitude of the primordial gravitational waves.

Before we proceed we need to clarify an important issue with regards to the term $\sim f_0\left (t-t_s
\right)^{\alpha}$ in Eq. (\ref{singstarobhub}). As it can be seen, since $H_i\gg f_0$, $M\gg f_0$ and also $f_0\ll 1$, this term is insignificant at early times, so practically it has no effect in the dynamics of the Hubble rate. However caution is needed at this point, since the behavior of the aforementioned term depends on the value of $\alpha$. For example, if $\alpha<-1$, then at $t=t_s$ a Big Rip singularity occurs, and this alters completely the behavior of the Hubble rate. But for soft singularities, no differences occur with the quasi de Sitter evolution $H(t)\sim H_i-\frac{M^2}{6}\left ( t-t_i\right )$. Then what is the motivation to use such a soft singularity extension. The answer to this very important question is that, although there is no difference at the Hubble rate level, there is significant difference at the physical quantities that contain higher derivatives of the Hubble rate, like for example the slow-roll parameters. Particularly, this study was performed in detail in Refs. \cite{sergeioikonomou1}, \\ \cite{sergeioikonomou2} and \\ \cite{sergeioikonomoujord}, where it was shown that even in the case of a Type IV singularity, the second slow-roll index may diverge at the singularity and only at that time instance, and this makes the dynamical evolution unstable at that time instance. Now the motivation for the present paper is to investigate whether the term $\sim f_0\left (t-t_s
\right)^{\alpha}$ has any effect on the amplitude of the gravitational waves. As we show in the next sections, only in the case of a Type IV singularity, the effect is insignificant. In all other finite-time singularities, non-trivial effects occur. This explains the motivation for using such a singular term in this paper, although the Hubble rate is unaffected, observables or physical quantities that contain higher derivatives of the Hubble rate, may diverge, depending on the singularity. The most intriguing case is the Type II case, as we show later on.

Let us now demonstrate in brief how to calculate the amplitude of the gravitational waves, by using the formalism and notation of Ref. \cite{oldie}. As was shown in \cite{oldie}, the metric of a wave with wavenumber $k$ has the following form,
\begin{equation}\label{metricgravwave}
\mathrm{d}s^2=-\mathrm{d}t^2+a(t)^2\left(\delta_{ij}+he_{ij} \right)\mathrm{d}x^i\mathrm{d}x^j\, ,
\end{equation}
with $e_{ij}$ being the polarization tensor which is symmetric, also it is traceless, that is $e^{i}_i=0$ and also satisfies the transverse condition $e_{ij}k^j=0$. The field equation for the gravitational wave, which is quantified in terms of the function $h$, for the $R^2$ gravity of Eq. (\ref{starobfr}) reads,
\begin{equation}\label{gravwaveini}
\ddot{h}+\left(3H+\frac{1}{36M^2}\frac{R^2}{(1+\frac{R}{3M^2})H}\right)\dot{h}-\frac{1}{a^2}\partial_i^2h=0\, .
\end{equation}
The quantization condition for the gravitational wave is,
\begin{equation}\label{quantcond}
[h(t,x),\dot{h}(t,x)]=i G\frac{\delta^3(x-y)}{a^3(1+\frac{R}{3M^2})}\, ,
\end{equation}
so the spacetime function $h(t,x)$ can be expanded in quantized wave modes as follows,
\begin{equation}\label{quantizedmmodes}
h(t,x)=\int\mathrm{d}k^3\left( a_k\,h_ke^{ikx}+a_k^{\dag}h_k^*e^{-ikx} \right)\, ,
\end{equation}
where $a_k$ and $a_k^{\dag}$ are the annihilation and creation operators which satisfy $[a_k,a_{k'}^{\dag}]=\delta^3(k-k')$, and also $h_k$ describes a gravitational wave mode with wavenumber $k$. By using the above relations, and also combining these with Eq. (\ref{gravwaveini}), the differential equation which governs the evolution of each gravitational wave mode $h_k$, is equal to,
\begin{equation}\label{gravifin}
\ddot{h}_k+\left(3H+\frac{1}{36M^2}\frac{R^2}{(1+\frac{R}{3M^2})H}\right)\dot{h}_k+\frac{k^2}{a^2}h_k=0\, .
\end{equation}
Note that the initial condition for the above equation is assumed to be a Bunch-Davies vacuum \\ \cite{bunch}. At early times, with $k>aH$, the wave was inside the horizon, so initially, $h_k$ is equal to,
\begin{equation}\label{initialhk}
h_k=h_k^ie^{-ik\int \frac{\mathrm{d}t}{a}}\, ,
\end{equation}
with $h_k^i$ being equal to,
\begin{equation}\label{difbr}
h_k^i=\frac{\sqrt{G}}{\sqrt{2k}(2\pi)^{3/2}a(1+\frac{R}{3M^2})^{1/2}}\, .
\end{equation}
As was shown in \cite{oldie}, when the horizon is crossed by the gravitational wave, say at the cosmic time instance $t_{hc}$, this gravitational wave mode reenters the horizon and becomes relevant for present time evolution. Therefore, the amplitude of the gravitational wave $A_k$ is \cite{oldie},
\begin{equation}\label{amplitudegravwave}
A_k=(2\pi k)^{3/2}|h_k(t_{hc})|=\frac{\sqrt{G}H(t_{hc})}{\sqrt{2k}(1+\frac{R(t_{hc})}{3M^2})^{1/2}}\, ,
\end{equation}
By using the above relation (\ref{amplitudegravwave}), we can directly see the effect of the finite-time singularities on the amplitudes of the gravitational waves. There are two cases we shall consider, firstly that the singularity time coincides with the horizon crossing time, and secondly if these two do not coincide. The second case is particularly interesting since the singularity time can be chosen to be at the end of the inflationary era.

The amplitude (\ref{amplitudegravwave}) for the Hubble rate (\ref{singstarobhub}) becomes,
\begin{equation}\label{amplcase1}
A_k=\frac{\sqrt{G} \left(H_i+\frac{1}{6} M^2 (-t_{hc}+t_i)+f_0 (t_{hc}-t_s)^{\alpha }\right)}{\sqrt{2} \sqrt{S(t)}}\, ,
\end{equation}
with the function $S(t)$ being equal to,
\begin{align}\label{st}
& S(t)=1+\frac{2 \left(-\frac{M^2}{6}+2 \left(H_i+\frac{1}{6} M^2 (-t_{hc}+t_i)\right) \right)}{M^2}\\ \notag & +\frac{2 \left(f_0 (t_{hc}-t_s)^{\alpha }\right)^2+f_0 (t_{hc}-t_s)^{-1+\alpha } \alpha }{M^2}\, .
\end{align}
Then, for a Type II singularity, since $0<\alpha<1$, the amplitude $A_k$ becomes approximately,
\begin{equation}\label{amplitudetypeII}
A_k\simeq \frac{\sqrt{G} M\left(H_i+\frac{1}{6} M^2 (-t+t_i)\right)\sqrt{(t_{hc}-t_s)^{1-\alpha }}}{\sqrt{2} \sqrt{f_0\text{  }\alpha }}\, .
\end{equation}
This means that in the case $t_{hc}=t_s$, the amplitude of the gravitational wave is $A_k\simeq 0$, and in the case that $t_s\neq t_{hc}$, the amplitude is proportional to $\sim (t_{hc}-t_s)^{1-\alpha}$, which means that $A_k$ is very small, since $t_{hc},t_s\ll 1$. This is a significant effect, and to some extent not so encouraging, since in the presence of a pressure singularity during inflation, or at the end of inflation, the amplitude of the primordial gravitational waves is almost zero, so this does not leave much possibility for the detection of the primordial wave modes.

In the case of the Type IV singularity, the results are most promising. In this case, since $\alpha>1$, the amplitude $A_k$ for $t_{hc}=t_s$,  reads,
\begin{equation}\label{aktypeiv}
A_k\simeq \frac{\sqrt{G} \left(H_i+\frac{1}{6} M^2 (-t_{hc}+t_i)\right)}{\sqrt{2} \sqrt{1+\frac{2 \left(-\frac{M^2}{6}+2 \left(H_i+\frac{1}{6} M^2 (-t_{hc}+t_i)\right)^2\right)}{M^2}}}\, ,
\end{equation}
which is identical to the predicted amplitude for the non-singular $R^2$ model. Hence in this case the Universe passes smoothly from the Type IV singularity, since even the primordial gravitational waves are not affected by the presence of this Type IV singularity. Moreover, even in the case $t_{hc}\neq t_s$, owing to the fact that the terms containing $\sim (t_{hc}-t_s)^{\alpha}$, and $\sim (t_{hc}-t_s)^{\alpha-1}$ are significantly smaller in comparison to the rest of the terms, the amplitude $A_k$ is approximately the same to the one appearing in Eq. (\ref{aktypeiv}). Therefore the Type IV singularity has practically no effect on the amplitude of the gravitational waves, and therefore in conclusion the Universe passes smoothly from a Type IV singularity.

\section{Conclusions}

In this paper we discussed the quantitative implications of finite-time singularities on the amplitude of the primordial gravitational waves. As we evinced, the only singularities that are allowed to occur during the inflationary era are the Type II, or so-called pressure singularity, and the Type IV singularity. With regards to the Type II singularity, in both the cases that the singularity occurs at the horizon crossing time instance, or at the end of the inflationary era time instance, the amplitude of the gravitational waves is zero or approximately zero. Hence, this is a big impact on the gravitational waves, however, the existence of a pressure singularity during inflation is not allowed to occur, as was discussed in \cite{sergeioikonomoujord}. In the case of a Type IV singularity, the amplitude of the gravitational waves modes is unaffected by the presence of the singularity, and therefore the Universe smoothly passes through the singularity without the catastrophic effects of crushing or even pressure singularities. Therefore the only effect of the Type IV singularities on the inflationary era is that it alters the dynamics of inflation and it can generate the graceful exit from inflation, as was shown in \cite{sergeioikonomoujord}.

In conclusion, the Type IV finite-time singularity has no effect on the gravitational waves, at least when $F(R)$ theories of gravity are considered. It would be interesting to discuss the same issue in the context of $F(G)$ theories of gravity, but we defer this task to a future work.

\section*{Acknowledgments}

Financial support by the Research
Committee of the Technological Education Institute of Serres,
under grant SAT/ME/220616-132/10, is gratefully acknowledged.

\end{document}